\documentclass{moriond}
\usepackage{amsmath,amsfonts,amssymb,latexsym,wasysym}

\def\Journal#1#2#3#4{{#1} {\bf #2}, #3 (#4)}

\def\PLB{{\em Phys. Lett.}  B}

\def\PRD{{\em Phys. Rev.} D}

\def\be{\begin{equation}}
\def\ee{\end{equation}}
\def\bea{\begin{eqnarray}}
\def\eea{\end{eqnarray}}

\def\ie{i.e.}
\def\eg{e.g.}

\newcommand{\Photo}{}

\begin{document}
\vspace*{4cm}
\title{Superradiance in deformed Kerr black holes}

\author{Mauro Oi$^{1,2}$, Edgardo Franzin$^{3,4,5}$, Stefano Liberati$^{3,4,5}$}

\address{\emph{$^1$}Università degli Studi di Cagliari, Dipartimento di Fisica, Cittadella Universitaria, 09042 Monserrato, Italy\\
\emph{$^2$}INFN, Sezione di Cagliari, Cittadella Universitaria, 09042 Monserrato, Italy\\
\emph{$^3$}SISSA, International School for Advanced Studies, via Bonomea 265, 34136 Trieste, Italy\\
\emph{$^4$}IFPU, Institute for Fundamental Physics of the Universe, via Beirut 2, 34014 Trieste, Italy\\
\emph{$^5$}INFN, Sezione di Trieste, via Valerio 2, 34127 Trieste, Italy}

\maketitle

\abstracts{
Ongoing observations in the strong-field regime are in optimal agreement with general relativity, although current errors still leave room for small deviations from Einstein's theory. Here we summarise our recent results on superradiance of scalar and electromagnetic test fields in Kerr-like spacetimes, focusing mainly on the Konoplya--Zhidenko metric. We observe that, while for large deformations with respect to the Kerr case superradiance is suppressed, it can be nonetheless enhanced for small deformations. We also study the superradiant instability caused by massive scalar fields, and we provide a first estimate of the effect of the deformation on the instability timescale.
}

\section{Introduction}
During its century of history, general relativity has been widely tested both in the weak and in the strong field regimes~\cite{LVTests,Will,Yunes,CP_review}. The recent detection of gravitational waves has further advanced the possibility to test deviations from the theoretical predictions and while present measurements are so far in agreement with Einstein's theory, current errors still leave room for small deviations. Indeed several modified theories of gravity and quantum gravity scenarios suggest that such deviations could be there, in particular in black hole spacetimes.

In this context, instead of testing the predictions of a specific model it is possible to describe discrepancies from the Kerr geometry by parametrising deviations from general relativity and trying to reconstruct the effects of such differences on observable quantities. Konoplya, Rezzolla and Zhidenko proposed a class of parametrised metrics to describe spherically or axisymmetric asymptotically flat black holes~\cite{RZ,KRZ} written as an expansion in the radial coordinate and the polar angle. Since it is compelling that astrophysical observable quantities depend only on few parameters, a minimal deformation from the Kerr metric have been proposed by Konoplya and Zhidenko~\cite{KZ_metric}. This Kerr-like spacetime preserves some the properties of the Kerr spacetime but, at the same time, it allows for substantial modifications of the black-hole structure at the horizon scale.

Rotating spacetimes exhibit an interesting behaviour when surrounded by perturbing fields. Consider the scattering of a wave with frequency $\omega$ and azimuthal number $m$ off a black hole: if the incident wave frequency satisfies the condition $\omega<m\Omega_H$, where $\Omega_H$ is the angular velocity on the horizon, the reflected wave gets its amplitude enhanced with respect to the incident one. This phenomenon is called black-hole superradiance~\cite{Zeldovich,BCP_review}. Moreover, in presence of a massive field, superradiant scattering can give rise to unstable and exponentially growing modes~\cite{sup_instability}, extracting more and more rotational energy from the black hole. Superradiant scattering and, in particular, the superradiant instability could be used \eg\ to constrain the mass of ultralight boson fields~\cite{ULbosons}.

In this work we present our recent results~\cite{FLO} on superradiance and superradiant instability in Konoplya--Zhidenko (KZ) black holes for test scalar and electromagnetic fields.

\section{The KZ black hole}
The KZ black hole is described, in Boyer--Lindquist-like coordinates, by the metric
\begin{align}\label{eq:KZ}
ds^2 =& %
-\left[1-\frac{2 M r^2 + \eta}{ r \left(r^2 + a^2 \cos ^2\theta\right)}\right]dt^2 - %
2 a \sin ^2\theta\frac{2 M r^2 + \eta}{r(r^2 + a^2 \cos ^2\theta)}dt\,d\varphi\nonumber\\ & + %
\frac{\left(r^2 + a^2\right)^2-a^2 \Delta  \sin ^2\theta}{r^2 + a^2 \cos ^2\theta}\sin ^2\theta d\phi^2 + %
\frac{r^2 + a^2 \cos ^2\theta}{\Delta }dr^2 + %
\left(r^2 + a^2 \cos ^2\theta\right)d\theta^2,
\end{align}
where $M$ is the ADM mass of the black hole, $a$ is the spin parameter, $\eta$ is some deformation parameter and $\Delta\equiv r^2-2Mr+a^2-\eta/r$. The metric~\eqref{eq:KZ} reduces to Kerr in the limit $\eta\to 0$.

For the KZ black hole the horizons are located at the roots of the equation $\Delta=0$, which admits up to three (possibly complex-valued) solutions, given by
\begin{equation}
r_k=\frac{2M}{3}+\frac{2}{3}\sqrt{4M^2-3a^2}\cos\left(\beta-\frac{2k\pi}{3}\right),\quad%
\beta=\frac{1}{3}\cos^{-1}\frac{16M^3-18Ma^2+27\eta}{2(4M^2-3a^2)^{3/2}},
\end{equation}
where $k=0,1,2$. Notice that, differently from the Kerr case, the parameter space is larger and no upper bounds on the spin parameter exist. Moreover, studying the spacetime structure, it is useful to introduce the quantity
\begin{equation}
\eta_-=\frac{2}{27}\left[9Ma^2-8M^3\-\left(4M^2-3a^2\right)^{3/2}\right].
\end{equation}
In this work we limit our analysis to the case $\eta>\eta_-$, in which the horizon is located at $r_0$, otherwise the spacetime could describe naked singularities.

For small values of the deformation parameter, \ie\ $\eta/M^3\ll 1$, and for $a<M$, the event horizon location is given by
\begin{equation}\label{eq:r0_smalleta}
r_0=r_++\frac{\eta}{r_+(r_+-r_-)}+O(\eta^2),
\end{equation}
where $r_+$ and $r_-$ are, respectively, the event and the Cauchy horizons of the Kerr black hole. From eq.~\eqref{eq:r0_smalleta} we interpret the deformation parameter as a displacement of the horizon with respect to $r_+$ and we see that the KZ black hole is less (more) compact than the Kerr spacetime with the same spin parameter for positive (negative) values of $\eta$.

The ergosurfaces are located at the roots of the equation $g_{tt}=r^2-2Mr+a^2\cos^2\theta-\eta/r=0$. For small deformations these are very similar to the Kerr's ones, however, for some specific values of the black-hole parameters, they can assume nonphysical discontinuous shapes~\cite{FLO}.

\section{Superradiant scattering from a KZ black hole}
We now consider scalar and electromagnetic perturbations propagating in a KZ background. Notice that, since the KZ metric is not a solution of a specific theory, we are not able to study gravitational perturbations and, although scalar and electromagnetic test fields are good proxies for studying black-hole properties, spin-2 perturbations could behave very differently.
It turns out that the Klein--Gordon ($s=0$) and Maxwell ($s=\pm 1$) equations separate into an angular part, described by the well known spin-weighted spheroidal harmonics, and a radial one, which satisfies a Teukolsky-like equation~\cite{Teukolsky}, given by
\begin{equation}\label{eq:pert}
\Delta^{-s}\frac{d}{dr}\left(\Delta^{s+1}\frac{dR_s}{dr}\right)+\left(\frac{K^2-2\text{i}s\Delta' K}{\Delta}+4\text{i}s r \omega-\lambda+\frac{s(s+1)(\Delta''-2)}{2}\right)R_s=0,
\end{equation}
where $K=(r^2+a^2)\omega-am$, $\omega$ is the wave frequency, $\lambda=A+a^2\omega^2-2am\omega$, being $A$ the eigenvalue of the angular equation, and a prime denotes differentiation with respect to the radial coordinate. As discussed in our work~\cite{FLO}, solutions of eq.~\eqref{eq:pert} with spin-weight $-s$ contain the same physical information as those with spin-weight $s$. For each spin-weight, the angular eigenvalue, and the corresponding radial solution, are also characterised by the harmonic number $l$ and the azimuthal number $m$.

Equation~\eqref{eq:pert} can be integrated when supplied by boundary conditions. In order to obtain such conditions, we introduce the tortoise radial coordinate $r_\ast$, defined by $dr_\ast/dr=(r^2+a^2)/\Delta$, and a new radial function $Y_s(r)=\sqrt{r^2+a^2}\Delta^{s/2}R_s(r)$. With this substitution, one finds that perturbations behave as $Y_s\sim r^{\pm s}e^{\mp\text{i}\omega r_\ast}$ at infinity, where the minus (plus) refers to ingoing (outgoing) waves, while at the horizon purely ingoing waves behave as $Y_s\sim\Delta^{-s/2}e^{-\text{i}kr_\ast}$, being $k=\omega-m\Omega_0$ and $\Omega_0=a/(2Mr_0+\eta/r_0)$ the angular velocity on the horizon.

Superradiant scattering occurs when an incident wave, with amplitude $\mathcal{I}$ and with frequency $\omega<m\Omega_0$, scatters off a black hole and its amplitude gets enhanced. Asymptotically, the radial function can be written as
\begin{equation}\label{eq:ansatz}
Y_s\sim \mathcal{I}e^{-\text{i}\omega r_\ast}r^s+\mathcal{R}e^{\text{i}\omega r_\ast}/r^s\quad \text{for } r_\ast\to\infty,
\end{equation}
where $\mathcal{R}$ is the reflected wave amplitude. We can then compute the ingoing and outgoing energy fluxes $dE_\text{in/out}/dt$, which are proportional to $|\mathcal{I}|^2$ and $|\mathcal{R}|^2$ respectively, and define the so-called amplification factors $Z_{s,l,m}=dE_\text{out}/dE_\text{in}-1$. $Z_{s,l,m}$ is positive when the outgoing flux is greater than the ingoing one meaning that energy is extracted from the black hole and superradiant scattering occurs.

\section{Results}
\begin{figure}
\centering
\includegraphics[scale=0.29]{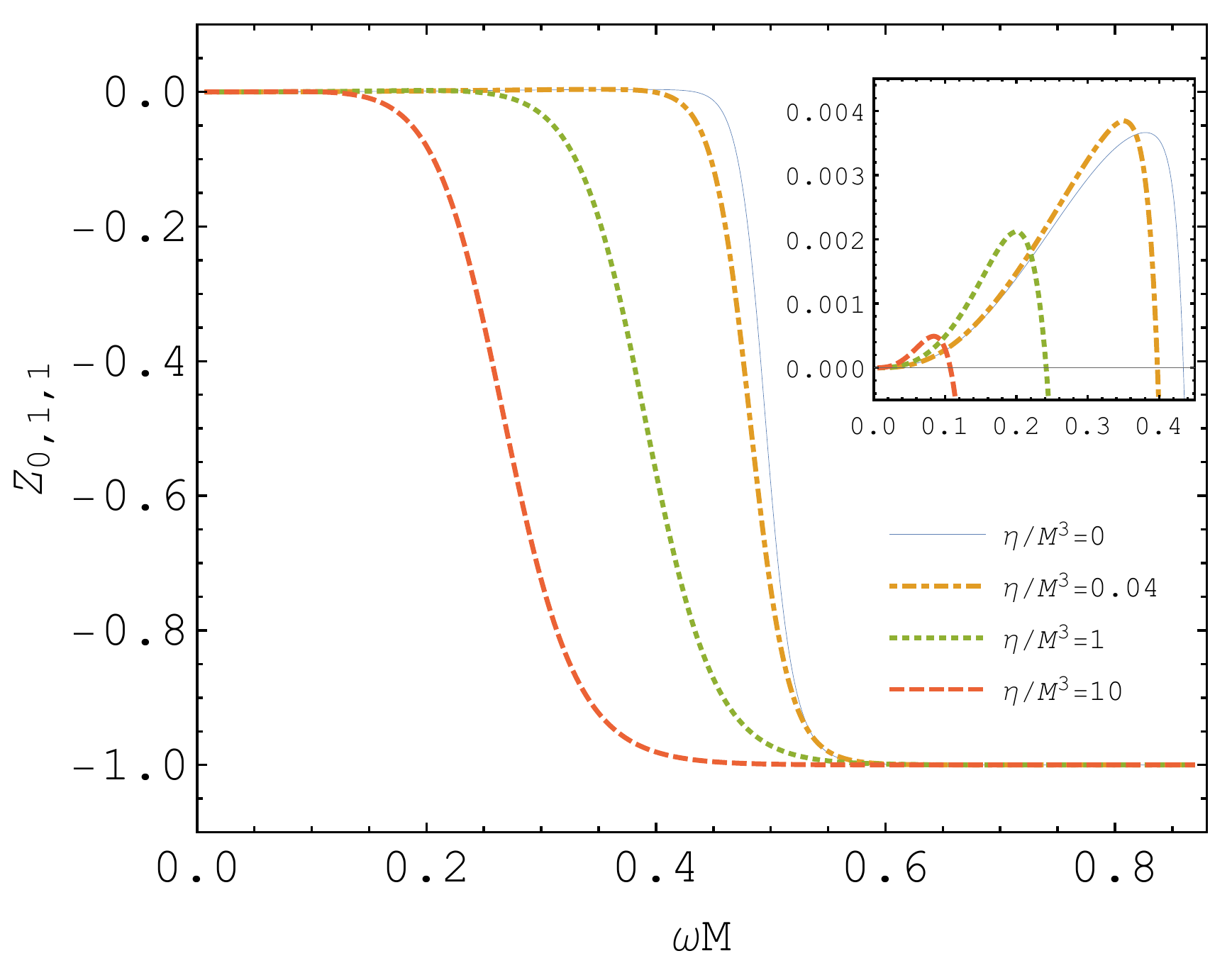}
\includegraphics[scale=0.29]{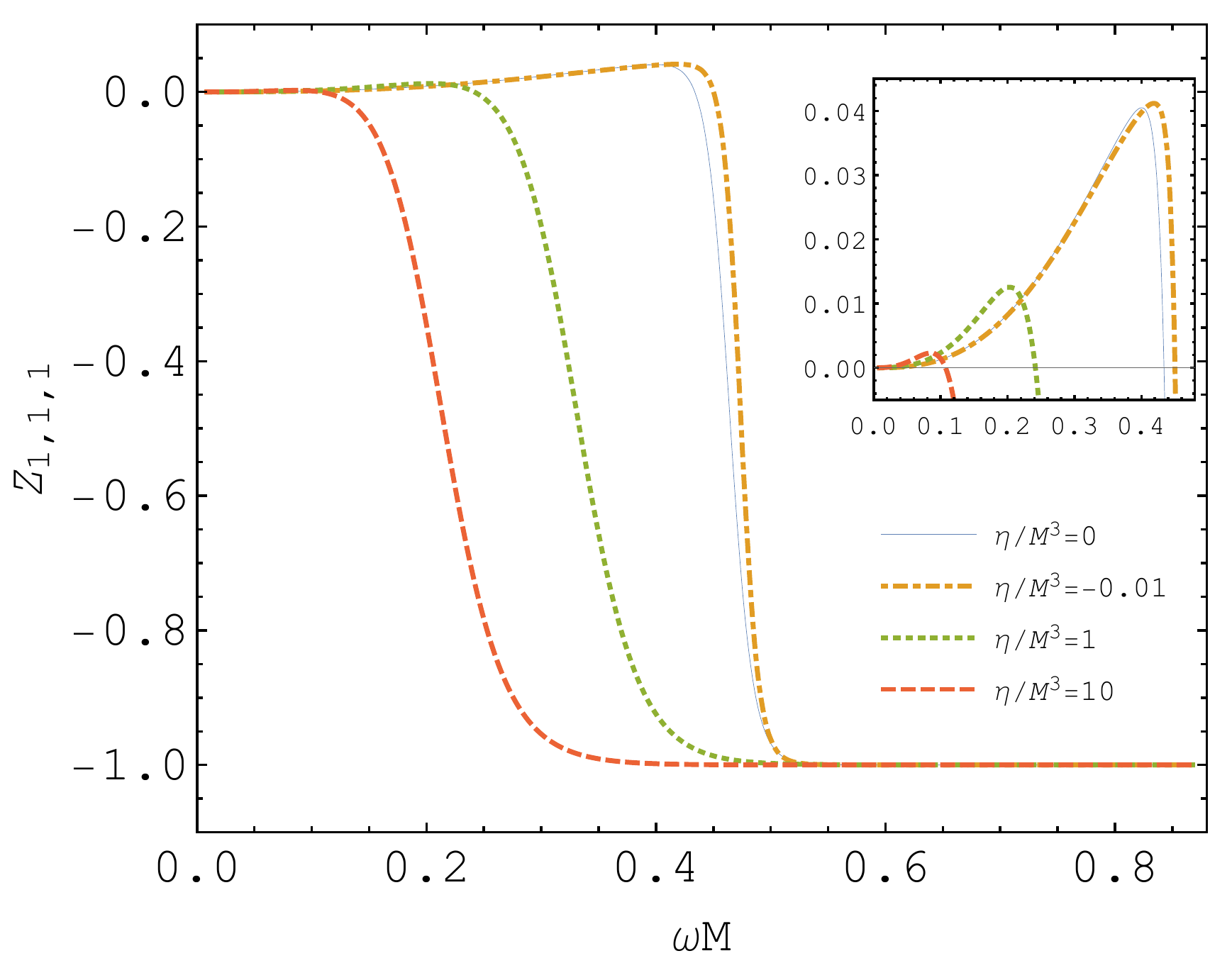}
\includegraphics[scale=0.29]{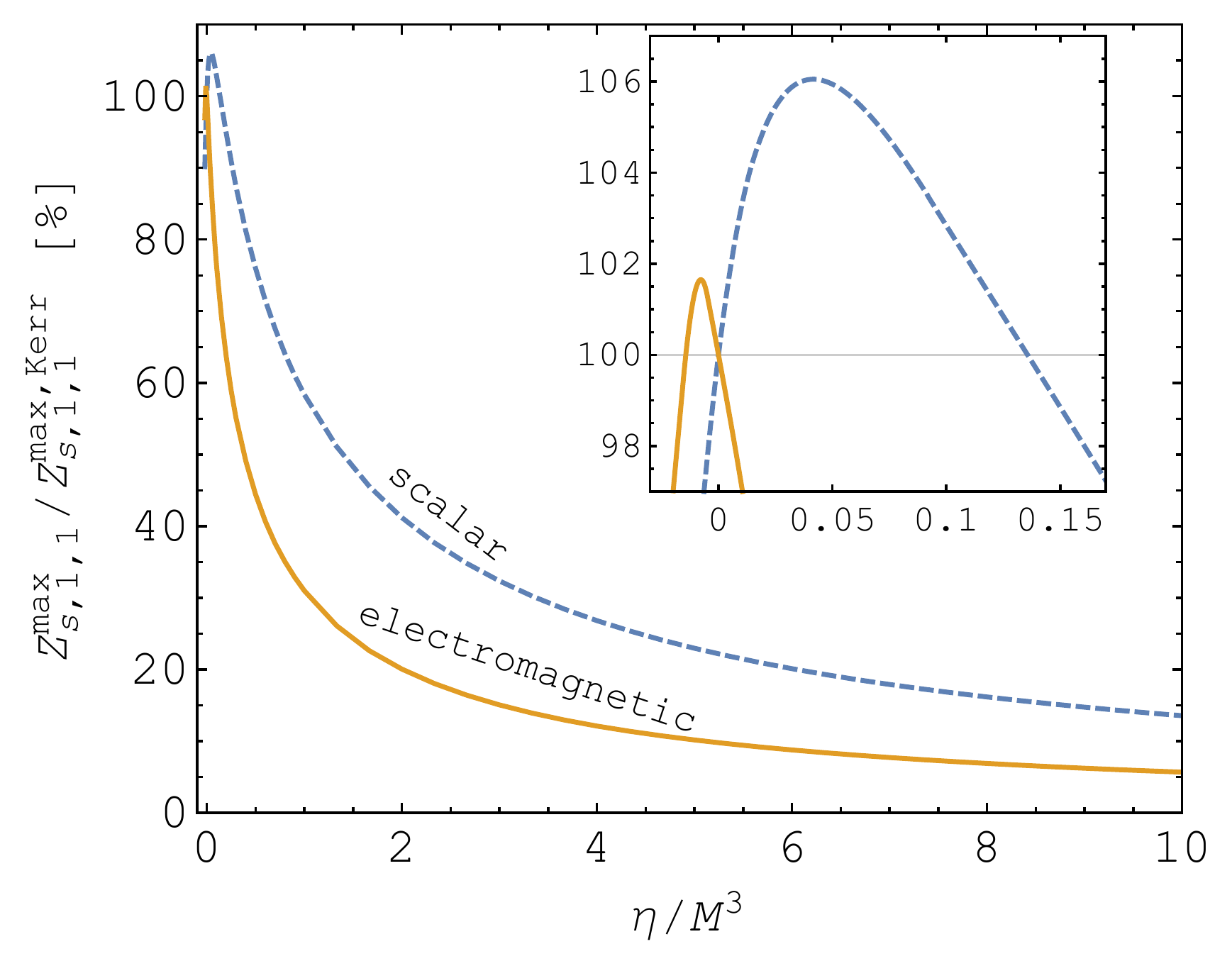}
\caption{\label{fig:spectra}Left and center panels: amplification factor spectra for $l=m=1$ scalar (left) and electromagnetic (center) modes around a KZ black hole with $a=0.99\,M$ and selected values of $\eta/M^3$. Right panel: amplification factor maximum as a function of the deformation parameter for $l=m=1$ and $a=0.99\,M$ normalised to the maximum value in Kerr.}
\end{figure}

In order to compute the amplification factors we solved eq.~\eqref{eq:pert} numerically for several values of the black-hole parameters and the field quantum numbers, as described in our work~\cite{FLO}. Some of our results for the dominant mode ($l=m=1$) are presented in fig.~\ref{fig:spectra}, but more can be found in our work and online~\cite{online}. Qualitatively, the amplification factors for a KZ black hole share the same behaviour with its Kerr analogous: in fact, $Z_{s,l,m}$ are positive only for $\omega<m\Omega_0$ and $m>0$, at $\omega\approx m\Omega_0$ the curve drops down very quickly and for $\omega\gg m\Omega_0$ $Z_{s,l,m}\simeq-1$, implying total absorption. Nonetheless, there are also some important differences. For large values of the deformation parameter, as $m\Omega_0$ becomes smaller, the integral of the amplification factor spectrum over the frequency range $(0, m\Omega_0)$ gets reduced and the superradiant scattering is suppressed. As in the Kerr case, the larger the spin parameter, the larger the amplification factors. Since the KZ black hole has no maximum value of $a$ above which the spacetime always describes a naked singularity, the amplification factors can, in principle, become very large as the black-hole spin increases. Surprisingly, we found that the maximum of $Z_{s,l,m}$ can be, for $a<M$ and small $\eta/M^3$, few percents greater than its Kerr analogous. For sufficiently large spin parameters ($a\gtrsim 0.97\,M$), this happens for positive values of the deformation parameter, \ie\ for black holes less compact than Kerr ones. In relative values, for $a\simeq0.99\,M$ and $\eta\simeq0.04\,M^3$, we observe a maximum amplification factor about $6\%$ and $1\%$ larger than the same quantity computed in Kerr for scalar and electromagnetic waves, respectively.

We also simulated the scattering of massive scalar waves with mass $m_s=\hbar\mu_s$. In this case, the boundary conditions at infinity get modified since modes with frequency $\omega<\mu_s$ are trapped near the horizon and asymptotically suppressed. Therefore, at infinity ($r_\ast\to\infty$ and $r\sim r_\ast$) we have $Y_0\sim r^{M(\mu_s^2-2\omega^2)/\tilde{\omega}}e^{\tilde{\omega}r}$, where $\tilde{\omega}=\pm(\mu_s^2-\omega^2)^{1/2}$. We limited our analysis to the dominant $l=m=1$ mode, for which we studied the amplification factors spectra and the superradiant instability. The spectra qualitatively behave as the massless ones, although for heavier perturbing fields the superradiant phenomenon gets suppressed. Regarding the instability, following Detweiler~\cite{Detweiler}, we showed that under the assumptions $M\omega\ll 1$, $|\eta|/M^3\ll 1$, $a\omega\ll 1$ and $M\mu_s\ll 1$ the problem can be treated analytically and we found that the growth time of the instability gets corrected by a quantity proportional to $\eta/M^3$. In particular, for an axion field with mass $m=10^{-20}\text{ eV}$ the instability time is given by
\begin{align}
\tau&=\tau_\text{Kerr}-(7.89\cdot 10^3\text{ s})\left(\frac{\mu_\text{axion}}{\mu_s}\right)\left(\frac{\eta/M^3}{0.01}\right)\left(\frac{M}{a}\right)^3\frac{1}{(M\mu_s)^8},
%
\end{align}
where $\tau_\text{Kerr}\simeq 10^6\text{ s}$ and from which we see that $\tau$ is shorter (longer) for positive (negative) values of the deformation parameter.

Finally, we found that for $\eta/M^3\ll 1$ massless scalar field in KZ and little massive fields in Kerr produce almost degenerate spectra.

\section*{Acknowledgements}
E.~F. and S.~L. acknowledge funding from the MIUR under the grant PRIN MIUR 2017-MB8AEZ. M.~O. acknowledges partial financial support by the research project CUP F71I17000150002, funded by Fondazione di Sardegna.

\end{document}